\documentclass[ %twocolumn,
aps,prd,   
               preprintnumbers,numbers,sort&compress,nofootinbib,
                            showpacs,
               colorlinks,
               linkcolor=blue,   
               citecolor=blue]{revtex4-1}
   \newcommand{\exclude}[1]{}

\usepackage{graphicx,amsmath,amssymb,bm}
\usepackage{psfrag}
\usepackage{feynmp}
\usepackage{hyperref}
\usepackage{enumitem}
\newcommand{\nn}{\nonumber}
\newcommand{\beq}{\begin{equation}}
\newcommand{\eeq}{\end{equation}}
\newcommand{\be}{\begin{eqnarray}}
\newcommand{\ee}{\end{eqnarray}}

\def\dd{ \,\mathrm{d} }

\def\+{\dagger}
\def\la{\langle}
\def\ra{\rangle}
\def\<{\langle}
\def\>{\rangle}

\newcommand{\Lqcd}{\Lambda_{\mathrm{QCD}}}

\begin{document}

\title{ QCD as  a topologically ordered system. }

\author{   Ariel R. Zhitnitsky} 
 \affiliation{Department of Physics \& Astronomy, University of British Columbia, Vancouver, B.C. V6T 1Z1, Canada}
%\date{\today}

\begin{abstract}
We argue that QCD belongs to a  topologically ordered phase similar to many well-known condensed matter systems with a gap such as topological insulators or superconductors. Our arguments are based on an analysis of the so-called ``deformed QCD" which is a weakly coupled gauge theory, but nevertheless  preserves all the crucial elements of strongly interacting QCD, including confinement, nontrivial $\theta$ dependence, degeneracy of the topological sectors,  etc.
 Specifically, we construct the so-called topological ``BF" action which reproduces the well known infrared features of the theory such as non-dispersive contribution to the topological susceptibility which can not be associated with any propagating degrees of freedom. Furthermore, we interpret the well known resolution of the celebrated  $U(1)_A$  problem where the would be $\eta'$ Goldstone boson generates its mass   as a result of mixing of the Goldstone field with a topological auxiliary field characterizing the system.  We then identify the non-propagating auxiliary topological field of the BF formulation in deformed QCD with the  Veneziano ghost (which plays  the crucial  role  in resolution of the  $U(1)_A$  problem). Finally, we elaborate on relation between ``string -net" condensation in topologically ordered condensed matter systems and long range coherent configurations, the ``skeletons",  studied in QCD lattice simulations. 
 
 \end{abstract}

\maketitle

\section{Introduction and motivation} \label{introduction}

The main motivation for the present studies is some recent lattice QCD  results   which can not be easily interpreted in terms of the conventional quantum field theory with a gap.   To be more specific, the gauge configurations studied in \cite{Horvath:2003yj,Horvath:2005rv,Horvath:2005cv,Alexandru:2005bn} display a laminar structure in the vacuum consisting of extended, thin, coherent, locally low-dimensional sheets of topological charge embedded in 4d space, with opposite sign sheets interleaved. A similar structure has also been 
observed in QCD by  different groups~\cite{Ilgenfritz:2007xu,Ilgenfritz:2008ia,Bruckmann:2011ve,Kovalenko:2004xm,Buividovich:2011cv,Blum:2009fh} and also in two dimensional $CP^{N-1}$ model~\cite{Ahmad:2005dr}.  The correlation length   of the percolating objects is order of size   of  the system   $\sim \mathbb{L}$ while the width of these objects  apparently vanishes in the continuum limit.   This is in drastic  contrast with the conventional expectation that in a gapped QCD in the hadronic phase 
 the fluctuations should have a typical scale $\sim \Lqcd$ while the gauge invariant correlation functions must display a conventional exponentially weak sensitivity to the size of the system $ \sim\exp(-\Lqcd{\mathbb  L })$.  Furthermore, the
studies of localization properties of Dirac eigenmodes have also shown evidence for the delocalization of low-lying modes onto effectively low-dimensional surfaces. 

It is important to emphasize that the observed long range configurations can not be identified with any   propagating degrees of freedom such as  physical gluons with transverse polarizations. In particular, these   long range objects contribute  to the contact term in topological susceptibility which has an on opposite   sign in comparison with conventional contributions related to   propagating physical states, see details below.  In other  words, lattice studies  are consistent   with    non-dispersive nature  of these long range objects.

It is not a goal of the present paper to cover this   subject \cite{Horvath:2003yj,Horvath:2005rv,Horvath:2005cv,Alexandru:2005bn,Ilgenfritz:2007xu,Ilgenfritz:2008ia,Bruckmann:2011ve,Kovalenko:2004xm,Buividovich:2011cv,Blum:2009fh,Ahmad:2005dr} with a number of subtle points which accompany it. 
In particular, all results \cite{Horvath:2003yj,Horvath:2005rv,Horvath:2005cv,Alexandru:2005bn,Ilgenfritz:2007xu,Ilgenfritz:2008ia,Bruckmann:2011ve,Kovalenko:2004xm,Buividovich:2011cv,Blum:2009fh,Ahmad:2005dr} which are sensitive to small   scales  (such as ``apparently vanishing width" or singular behaviour of the topological susceptibility at small $x$, see below)  should be taken with  great caution because of high sensitivity to the lattice size. However, 
there are many technical tools to ensure  that the obtained results are not some artifacts of the state of the art lattice simulations, but rather, represent  genuine physical results which properly describe the continuum limit\footnote{In particular, the question about small $x$ behaviour has been specifically addressed in  \cite{Ilgenfritz:2007xu} where it has been 
found that a singular behaviour at arbitrary small $x$ is a genuine physical result, rather than a lattice  artifact.}.

Our goal is in fact quite different. We attempt to  answer the following question: how could  the long range structure (if it is confirmed by future numerical   computations)   ever emerge in a gapped theory such as QCD? Our strategy to address this question is   to use the  ``deformed QCD'' as a toy model \cite {Shifman:2008ja,Yaffe:2008}. We do not even attempt to make precise comparison with lattice results \cite{Horvath:2003yj,Horvath:2005rv,Horvath:2005cv,Alexandru:2005bn,Ilgenfritz:2007xu,Ilgenfritz:2008ia,Bruckmann:2011ve,Kovalenko:2004xm,Buividovich:2011cv,Blum:2009fh,Ahmad:2005dr} in the present work. Our goal is much more modest;  we want to see if  such a  long range structure, in any form, could  emerge  in ``deformed QCD''.   This is a simplified  version of QCD which, on one hand, is  a weakly coupled gauge theory wherein computations can be performed in theoretically controllable manner. On other hand, the corresponding  deformation  preserves all the relevant elements of strongly coupled QCD such as confinement, degeneracy of topological sectors, nontrivial $\theta$ dependence, and many other important  aspects which allow us to test some fascinating   features of strongly interacting QCD, including some aspects of the  long range order. It is claimed \cite {Shifman:2008ja,Yaffe:2008} that there is no phase transition in passage from weakly coupled ``deformed QCD" to strongly coupled real QCD.  The ground state in ``deformed QCD"   is saturated by the fractionally charged weakly interacting pseudo-particles (monopoles) which live in 3 dimensions. 
 
 That the model is in 3d is precisely the feature which  offers a new  perspective    on a possible deep   relation  between  ``deformed QCD"  and  previously  studied   $(2+1)$ condensed matter (CM) systems with  a gap which   are known to lie in topologically 
 ordered phases~\cite{Wen:1989iv}. 
 These CM systems include, but are not limited to   such systems as: quantum hall states as described in  \cite{Wen:1989iv,Wen:1990zza,Moore:1991ks}, superconductors as presented in \cite{BF}, and topological insulators as treated in \cite{Cho:2010rk}.  See the recent review  paper \cite{Wen:2012hm} with large number of original references   on analysis  of the topological order in condensed matter systems. 
 The crucial  element which is common for all these CM systems is that they can be formally described using a pure topological Chern-Simons effective Lagrangian written in the so-called ``BF" form. 
 
 One of the main objectives  of this work  is to derive  a similar Chern-Simons theory 
 for the ``deformed QCD'' to argue  that it   also lies in  topologically ordered phase. 
 Of course, there is a fundamental difference between   CM systems defined in Minkowski space  and Euclidean 3d 
``deformed QCD". In particular, instead of propagating quasiparticles in CM systems we have pseudoparticles which saturate the path integral. 
Furthermore, the ``degeneracy"  in ``deformed QCD''  
  is related to degenerate winding states $| n\ra$ which are connected to each other by large gauge transformation, and therefore
 must be identified as they  correspond to   the same physical state.
     It is very different from  conventional degeneracy in  topologically ordered CM systems  when {\it distinct} degenerate states are present in the system. 
  Nevertheless,   the systems in both cases demonstrate a   sensitivity to the large distance physics formulated in terms of the boundary conditions, see few  additional comments in \cite{Zhitnitsky:2011aa}
on  analogies between topologically ordered CM systems and long range features in QCD.   
  In what follows we discuss some  manifestations  of this long range order in QCD which may be considered as   a complementary sign, along
with well-established characteristics such as topological entanglement entropy \cite{Kitaev:2005dm,Levin:2006zz}.
 
We start our presentation  in section~\ref{deformedqcd}  by reviewing  the relevant parts of the model \cite {Shifman:2008ja,Yaffe:2008}. In section~\ref{BF-section} we argue that the   infrared description can be formulated in terms of  Chern-Simons effective Lagrangian in the BF form. We reproduce the expression for the topological susceptibility using this BF-type action. In section \ref{analogy} we present the physical interpretation of the obtained results in   analogy 
with CM systems.    In section~\ref{conclusion} we speculate on  some   profound consequences 
 of  the topological order if it persists in   a passage  from weakly coupled ``deformed QCD" to strongly coupled physical QCD. 
     
\section{Deformed QCD} \label{deformedqcd}

Here we overview  the ``center-stablized" deformed Yang-Mills developed in \cite {Shifman:2008ja,Yaffe:2008}.  In the deformed theory an extra term is put into the Lagrangian in order to prevent the center symmetry breaking that characterizes the QCD phase transition between ``confined" hadronic matter and ``deconfined" quark-gluon plasma. The nature of  the gap  in this model     is reviewed in section~\ref{model}, while   in section \ref{top} we review  the computation  of the non-dispersive contact term in topological susceptibility \cite{Thomas:2011ee}. This term will be our starting point in construction of the corresponding Chern Simons Lagrangian in section \ref{BF-section}.
 
\subsection{The model}\label{model}

We start with pure Yang-Mills (gluodynamics) with gauge group $SU(N)$ on the manifold $\mathbb{R}^{3} \times S^{1}$ with the standard action
\be \label{standardYM}
	S^{YM} = \int_{\mathbb{R}^{3} \times S^{1}} d^{4}x\; \frac{1}{2 g^2} \mathrm{tr} \left[ F_{\mu\nu}^{2} (x) \right],
\ee
and add to it a deformation action,
\be \label{deformation}
	\Delta S \equiv \int_{\mathbb{R}^{3}}d^{3}x \; \frac{1}{L^{3}} P \left[ \Omega(\mathbf{x}) \right],
\ee 
built out of the Wilson loop (Polyakov loop) wrapping the compact dimension
\be \label{loop}
	\Omega(\mathbf{x}) \equiv \mathcal{P} \left[ e^{i \oint dx_{4} \; A_{4} (\mathbf{x},x_{4})} \right].
\ee
The parameter  $L$ here  is the length of the compactified dimension
which is assumed to be small. 
 The coefficients of the polynomial  $ P \left[ \Omega(\mathbf{x}) \right]$ can be suitably chosen such that the deformation potential (\ref{deformation}) forces unbroken symmetry at any compactification scales. At small compactification $L$ the gauge coupling  is small so that 
the semiclassical computations are under complete theoretical control \cite {Shifman:2008ja,Yaffe:2008}.

As described in \cite {Shifman:2008ja,Yaffe:2008}, the proper infrared description of the theory is a dilute gas of $N$ types of monopoles, characterized by their magnetic charges, which are proportional to the simple roots and affine root $\alpha_{a} \in \Delta_{\mathrm{aff}}$ of the Lie algebra for the gauge group $U(1)^{N}$. 
 For a fundamental monopole with magnetic charge $\alpha_{a} \in \Delta_{\mathrm{aff}}$, the topological charge is given by
\be \label{topologicalcharge}
	Q = \int_{\mathbb{R}^{3} \times S^{1}} d^{4}x \; \frac{1}{16 \pi^{2}} \mathrm{tr} \left[ F_{\mu\nu} \tilde{F}^{\mu\nu} \right]
		= \pm\frac{1}{N},
\ee
and the Yang-Mills action is given by
\be \label{YMaction}
	S_{YM} = \int_{\mathbb{R}^{3} \times S^{1}} d^{4}x \; \frac{1}{2 g^{2}} \mathrm{tr} \left[ F_{\mu\nu}^{2} \right]= \frac{8 \pi^{2}}{g^{2}} \left| Q \right|.
		 \ee
 The  $\theta$-parameter in the Yang-Mills action can be included in conventional way,
\be \label{thetaincluded}
	S_{\mathrm{YM}} \rightarrow S_{\mathrm{YM}} + i \theta \int_{\mathbb{R}^{3} \times S^{1}} d^{4}x\frac{1}{16 \pi^{2}} \mathrm{tr}
		\left[ F_{\mu\nu} \tilde{F}^{\mu\nu} \right],
\ee
with $\tilde{F}^{\mu\nu} \equiv \epsilon^{\mu\nu\rho\sigma} F_{\rho\sigma}$.

The system of interacting monopoles, including $\theta$ parameter, can be represented in the dual sine-Gordon form as follows \cite{Shifman:2008ja,Yaffe:2008,Thomas:2011ee},
\be
\label{thetaaction}
	S_{\mathrm{dual}}&=& \int_{\mathbb{R}^{3}}  d^{3}x \frac{1}{2 L} \left( \frac{g}{2 \pi} \right)^{2}
		\left( \nabla \bm{\sigma} \right)^{2} \nonumber \\&-& \zeta  \int_{\mathbb{R}^{3}}  d^{3}x \sum_{a = 1}^{N} \cos \left( \alpha_{a} \cdot \bm{\sigma}
		+ \frac{\theta}{N} \right)  	, 
\ee
where $\zeta$ is magnetic monopole fugacity which can be explicitly computed in this model using the conventional semiclassical approximation. The $\theta$ parameter enters the effective Lagrangian (\ref{thetaaction}) as $\theta/N$ which is the direct consequence of the fractional topological charges of the monopoles (\ref{topologicalcharge}). Nevertheless, the 
theory is still $2\pi$ periodic. This
  $2\pi$ periodicity of the theory is restored not due to the $2\pi$ periodicity of Lagrangian (\ref{thetaaction}).
  Rather, it is restored as a result of   summation over all branches of the theory when the  levels cross at
   $\theta=\pi (mod ~2\pi)$ and one branch replaces another and becomes the lowest energy state as discussed in \cite{Thomas:2011ee}.
 
Finally, the dimensional parameter which governs the dynamics  of the problem is the Debye   correlation length of the monopole's gas, 
 \be \label{sigmamass}
	m_{\sigma}^{2} \equiv L \zeta \left( \frac{4\pi}{g} \right)^{2}.
\ee
   The average number of monopoles in a ``Debye volume" is given by
\begin{equation} \label{debye}
{\cal{N}}\equiv	m_{\sigma}^{-3} \zeta = \left( \frac{g}{4\pi} \right)^{3} \frac{1}{\sqrt{L^3 \zeta}} \gg 1,
\end{equation} 
The last inequality holds since the monopole fugacity is exponentially suppressed, $\zeta \sim e^{-1/g^2}$, and in fact we can view (\ref{debye}) as a constraint on the validity of the  approximation where semiclassical approximation is justified. 
 
 \subsection{Topological susceptibility}\label{top}
 The topological susceptibility $\chi$
 which plays a crucial role in resolution of the $U(1)_A$ problem~\cite{witten,ven,vendiv,Rosenzweig:1979ay,Nath:1979ik,Kawarabayashi:1980dp}  is defined as follows\footnote{We use the Euclidean notations  where  path integral computations are normally performed.}
\be
\label{chi}
 \chi (\theta &=&0) =    \left. \frac{\partial^2E_{\mathrm{vac}}(\theta)}{\partial \theta^2} \right|_{\theta=0}  \\&=& \lim_{k\rightarrow 0} \int \!\dd^4x e^{ikx} \la T\{q(x), q(0)\}\ra , \nonumber
 \ee
where  $\theta$ is the  $\theta$ parameter which enters the  Lagrangian (\ref{thetaincluded}) along with  topological density operator $q (x)$ and $E_{\mathrm{vac}}(\theta)$ is the vacuum energy density determined by 
(\ref{thetaaction}).

It is important  that the topological susceptibility $\chi$  does not vanish in spite of the fact that $q= \partial_{\mu}K^{\mu}$ is total divergence.  Furthermore, any physical   state gives a negative contribution to this 
diagonal correlation function
\be	\label{G}
  \chi_{\rm dispersive} &\sim&  \lim_{k\rightarrow 0} \int d^4x e^{ikx} \la T\{q(x), q(0)\}\ra  \\
  &\sim & 
    \lim_{k\rightarrow 0}  \sum_n \frac{\la  0 |q|n\ra \la n| q| 0\ra }{-k^2-m_n^2}\simeq -\sum_n\frac{|c_n|^2}{m_n^2} \leq 0, \nonumber
\ee
 where   $m_n$ is the mass of a physical state,  $k\rightarrow 0$  is  its momentum, and $\la 0| q| n\ra= c_n$ is its coupling to topological density operator $q (x)$.
 At the same time the resolution of the $U(1)_A$ problem requires a positive sign for the topological susceptibility (\ref{top1}), see the original reference~\cite{vendiv} for a thorough discussion, 
\be	\label{top1}
  \chi_{\rm non-dispersive}= \lim_{k\rightarrow 0} \int \!\dd^4x e^{ikx} \la T\{q(x), q(0)\}\ra > 0.~~~
\ee
Therefore, there must be a contact contribution to $\chi$, which is not related to any propagating  physical degrees of freedom,  and it must have the ``wrong" sign. The ``wrong" sign in this paper implies a sign 
  which is opposite to any contributions related to the  physical propagating degrees of freedom (\ref{G}).   In the framework \cite{witten} the contact term with ``wrong" sign  has been simply postulated, while in refs.\cite{ven,vendiv} the Veneziano ghost had been introduced into the theory to saturate the required property (\ref{top1}).   
  Furthermore, as we discuss below the  contact term has  the structure $\chi \sim \int d^4x \delta^4 (x)$.
  The significance of this structure is  that the gauge variant correlation function in momentum space
  \be
  \label{K}
   \lim_{k\rightarrow 0} \int d^4x e^{ikx} \la K_{\mu}(x) , K_{\nu}(0)\ra\sim   \frac{k_{\mu}k_{\nu}}{k^4}
   \ee 
  develops  a topologically protected  ``unphysical" pole. Furthermore, the residue of this   pole has the ``wrong sign", which  precisely corresponds to the Veneziano ghost  contribution saturating the non-dispersive term  in gauge invariant correlation function (\ref{top1}),
   \be
  \label{K1}
   \< q({x}) q({0}) \> \sim  \la \partial_{\mu}K^{\mu}(x) , \partial_{\mu}K^{\mu}(0)\ra \sim \delta^4(x)
   \ee 
   The singular behaviour of
  $\< q(\bold{x}) q(\bold{0}) \>$   with the ``wrong sign" has been well confirmed by the lattice simulations \cite{Horvath:2005cv,Ilgenfritz:2007xu,Ilgenfritz:2008ia,Bruckmann:2011ve,lattice}.

 The topological susceptibility in the ``deformed QCD"  model 
 can be explicitly computed as this model is a weakly coupled gauge theory. In this model it is saturated by fractionally charged weakly interacting monopoles, and it is given by \cite{Thomas:2011ee}
\be \label{YM}
  \chi_{YM} = \int d^4 x \< q(\bold{x}) , q(\bold{0}) \>	=\frac{\zeta}{NL} \int d^3 x \left[ \delta(\bold{x}) \right].
\ee
It has the required  ``wrong sign" as this  contribution is not related to any physical propagating degrees of freedom, and it has a 
$\delta(\bold{x})$ function structure which implies the presence of the pole (\ref{K}). However, there are not any physical massless states in the system, and  the computations \cite{Thomas:2011ee} leading to (\ref{YM}) are accomplished without any ghosts or any other unphysical degrees of freedom. 
Instead, this term is described in terms of the tunnelling events between  different (but physically equivalent) topological sectors in the system. 

 One should emphasize that  the $\delta (\mathbf{x})$ function which appears in the expression for topological susceptibility (\ref{YM}) is not an artifact of small size monopole-
 approximation used in \cite{Thomas:2011ee}. Instead, this singular behaviour is a generic feature which is shared by many other models, including the exactly solvable two dimensional Schwinger model and also QCD with the contact term  saturated by the Veneziano ghost. In fact, 
 this singular behaviour is  measured in the QCD lattice simulations at strong coupling \cite{Horvath:2005cv,Ilgenfritz:2007xu,Ilgenfritz:2008ia,Bruckmann:2011ve,lattice} as we already mentioned.  One should emphasize again that the significance of this structure is  that the gauge variant correlation function in momentum space develops  a topologically protected  ``unphysical" pole similar to eq. (\ref{K}) which will play an important role in our discussions in section \ref{BF-section} 
 wherein  this pole will be identified with unphysical topological fields in the BF-formulation of the theory. 
 
 The $\delta (\mathbf{x})$ function in  (\ref{YM}) should be understood as total divergence related to the infrared (IR) physics, rather than to ultraviolet (UV) behaviour  as explained in~\cite{Thomas:2011ee}
\be	\label{divergence}
 &&\chi_{YM}\, \sim   \int \delta (\mathbf{x})  \dd^3x \nonumber\\
&=&  \int   \dd^3x~
  \partial_{\mu}\left(\frac{x^{\mu}}{4\pi x^3}\right)=
    \oint_{S_2}    \dd\Sigma_{\mu}
 \left(\frac{x^{\mu}}{4\pi x^3}\right).
\ee
 In other words, the non-dispersive contact term with the ``wrong" sign (\ref{YM}) is {\it highly sensitive} to the boundary conditions and behaviour of the system at arbitrarily  large distances. Therefore, it is natural to expect that a variation of the boundary conditions would change the topological susceptibility (\ref{YM}) despite of the fact that the system has a gap (\ref{sigmamass}). 
 We will  reproduce the $\delta (\mathbf{x})$ function in  (\ref{YM}) 
 in terms of the  topological quantum field theory for deformed QCD constructed in the next section \ref{top_bf}. This will further illuminate the IR nature of the contact term.

The light quarks in the fundamental representation can be easily inserted into the system \cite{Thomas:2011ee}.
 In the dual sine-Gordon theory   the $\eta'$ field  (represented in what follows  by dimensionless  $\phi$ field)  appears  exclusively in combination with the $\theta$ parameter as $\theta \rightarrow \theta - \phi$ as a consequence of the Ward Identities. Indeed, the transformation properties of the path integral measure under the chiral transformations $\psi\rightarrow \exp(i \gamma_5\frac{\phi}{2})\psi$ dictate  that $\eta'$ appears only in the combination  $\theta \rightarrow \theta - \phi$.  Therefore we have,
\be
\label{matter}
	S_{\mathrm{dual}}&=& \int_{\mathbb{R}^{3}}  d^{3}x \left[ \frac{1}{2 L} \left( \frac{g}{2 \pi} \right)^{2}
		 \left( \nabla \bm{\sigma} \right)^{2} + \frac{c}{2} \left( \nabla \phi \right)^{2}\right] \nonumber \\&-& \zeta  \int_{\mathbb{R}^{3}}  d^{3}x \sum_{a = 1}^{N} \cos \left( \alpha_{a} \cdot \bm{\sigma}
		+ \frac{\theta-\phi}{N} \right)  	, 
\ee
where   numerical coefficient $c/L$  has dimension $(mass)^2$   and plays the role of $f_{\eta'}^2$ in this model. The parameter $c$  can, in principle, be computed\footnote{Such a computations would require the calculation of  the chiral condensate as a first step.  After that, the $\eta'$   field should be identified with the phase of this condensate.  Fortunately, the outcome of this ``would be" calculation is known exactly  (\ref{matter}) as a consequence of the Ward Identities. }.  
However, such a computation is  beyond  the scope of the present studies.  
We   express our results  in terms of the $\eta'$ mass which  assumes the following value
\be
\label{mass}
m_{\eta'}^{2} = \frac{  \zeta}{cN}  =    \chi_{YM}\cdot  \frac{L }{c}.
\ee
Generation of the $\eta'$ mass proportional to $\chi_{YM}$ computed in pure YM theory  
represents a  precise realization of Witten's -Veneziano resolution of the $U(1)_A$ problem \cite{witten,ven,vendiv,Rosenzweig:1979ay,Nath:1979ik,Kawarabayashi:1980dp}.  
The crucial point here is that  the mass gap for the $ \bm{\sigma}$ field and for the $\phi$ field are generated by one and the same $\theta$ dependent potential (\ref{matter}) with one and the same coupling $\zeta$. This is because the   $\phi$ field  corresponding to the physical $\eta'$ field enters the effective potential in unique and precise form  $(\theta - \phi)$ as a consequence of the Ward Identities. An additional factor $1/N$ which appears for $m_{\eta'}^{2}$ (in comparison with $ m_{\sigma}^2$) is a precise realization of the original idea that $\theta$ parameter enters eq. (\ref{matter}) as $\theta/N$.

The corresponding generalization of eq. (\ref{YM}) for the topological susceptibility reads \cite{Thomas:2011ee}
\be \label{chi_QCD}
  \chi_{QCD} &=& \int d^4 x \< q(\bold{x}) q(\bold{0}) \>  \\
    &=&\frac{\zeta}{ NL} \int d^3 x \left[ \delta(\bold{x})-m_{\eta'}^2 \frac{e^{-m_{\eta'}r}}{4\pi r}  \right].  \nn
\ee
The first term in this equation has a  non-dispersive nature  and   has the positive sign. 
As explained above this contact term is not related to any physical propagating degrees of freedom.  Instead,  it emerges as a result of the tunnelling    transitions between the degenerate topological sectors in pure YM theory\footnote{\label{degeneracy}
  In the context of this paper the ``degeneracy'' implies there existence of winding states $| n\ra$ constructed as follows: ${\cal T} | n\ra= | n+1\ra$.  In this formula the operator ${\cal T}$ is  the  large gauge transformation operator  which commutes  with the Hamiltonian $[{\cal T}, H]=0$. The physical vacuum state is {\it unique} and constructed as a superposition of $| n\ra$ states. In QFT approach the presence of $n$ different sectors in the system is reflected  by  summation over $ { n \in \mathbb{Z}}$ in  definition of the path integral in QCD. It should not be confused with conventional term ``degeneracy'' when two or more physically {\it distinct} states are present in the system.
}. 
The positive sign of this term is the crucial element for the resolution of the $U(1)_A$ problem. The second term in eq. (\ref{chi_QCD}) is a standard dispersive  contribution, can be restored through the absorptive part using conventional dispersion relations, and has a negative sign  in accordance with general principles (\ref{G}). This conventional physical contribution is saturated in this model by the lightest $\eta'$ field. It enters $\chi_{QCD}$ precisely in such a way that the Ward Identity (WI) expressed as $\chi_{QCD} (m_q=0)=0$ is automatically satisfied as a result of cancellation between the two terms. If the non-dispersive contact  term with the ``wrong sign'' was not present in the system, the WI could not be satisfied as physical states always contribute with a negative sign in eqs. (\ref{G},\ref{chi_QCD}). 
 
  \section{BF theory for deformed QCD}\label{BF-section}
  In many respects the deformed QCD which is an effective  3d theory defined by the  action (\ref{thetaaction})  is very similar to the abelian Higgs model  describing superconductivity. The Higgs  theory is known to belong to the  topologically ordered phase \cite{BF}. Therefore, our goal here is to provide some arguments suggesting that these two models   in fact behave very similarly in the infrared regime. This similarity suggests that the deformed QCD also lies in a topological phase. Separately, it has been claimed  \cite {Shifman:2008ja,Yaffe:2008} that the passage from the deformed QCD to strongly coupled QCD  is smooth, without any phase transitions. If true, this would imply that QCD also lies in a topologically ordered phase. 
    
  \subsection{BF theory for the deformed QCD. Construction.}\label{BF}
  The action (\ref{thetaaction}) describes the theory with a gap, similar to the conventional Landau-Ginsburg effective action describing the  abelian Higgs model. In both cases the crucial part related to the topological order is missing in these effective actions. We want to reconstruct   this missing   topological term.  We follow \cite{BF} in this derivation. As our 3d model  (\ref{thetaaction}) has the Euclidean metric, we proceed with the Euclidean path integral  computations. It is different from the physical case of the Minkowski 2+1 BF theory discussed   in \cite{BF} in which  one can discuss the braiding phases of quasiparticles, the Hamiltonian formulation, etc.   Nevertheless, the crucial points   can be explained using the Euclidean path integral approach.

  We wish to derive the topological action for deformed QCD by using the same  technique exploited   in \cite{BF} for the Higgs model. 
  We start with the construction of the source term  for a configuration of $M^{(a)}$ point-like magnetic monopoles coupled to $\bm{\sigma}$ field   entering the low energy action  (\ref{thetaaction}) 
  \be 
  \label{source_m}
  {\cal{S}}_J &=&-\beta \int_{\mathbb{R}^{3}} d^{3}x J(\mathbf{x}) \bm{\sigma}  \\
	J(\mathbf{x}) &=&\frac {i}{\beta} \sum_{a = 1}^{N} \sum_{k = 1}^{M^{(a)}} Q_{k}^{(a)}
		\delta (\mathbf{x}_{k}^{(a)} - \mathbf{x}) \alpha_{a}, \nonumber
\ee
  where $ Q_{k}^{(a)}=\pm 1$, and  $\alpha_{a} \in \Delta_{\mathrm{aff}}$ is the affine root  of the Lie algebra for the gauge group $U(1)^{N}$ and beta is defined as 
\begin{equation} 
\label{beta}
	\beta \equiv \frac{1}{L} \left( \frac{g}{2\pi} \right)^{2}.
\end{equation}
It has been demonstrated in  \cite {Shifman:2008ja,Yaffe:2008} that summation over all possible positions and orientations of the monopoles leads to the low energy action (\ref{thetaaction}), see also \cite{Thomas:2011ee} with some technical details.

The   $\bm{\sigma}$ field    plays the role of the scalar magnetic potential as one can see from expression for the $U(1)^N$  magnetic field  $B^i = \epsilon^{ijk4} F_{jk}/2g$ 
\be \label{magnetic}
  F_{ij}^{(a)}=\frac{g^2}{2\pi L}\epsilon_{ijk}\partial^k \sigma^{(a)}, 
    ~~~ \mathbf{B}^{(a)}=\frac{g}{2\pi L}\nabla  \sigma^{(a)}.
\ee
Essentially, the $ \sigma^{(a)}$ fields are the physical photons in effectively three dimensional  space.   They become massive as a result of the Debye screening which eventually determines their masses (\ref{sigmamass}). These fields are dynamical fields, and  the corresponding Maxwell term $\frac{1}{2}(\mathbf{B}^{(a)}\cdot \mathbf{B}^{(a)})$ is expressed in eq. (\ref{thetaaction}) in terms of the $ \sigma^{(a)}$ fields.

  We now turn to an analysis of the   topological density operator $q(\mathbf{x})$ computed for background monopole configurations. It can be represented     in terms of the effective scalar $\bm{\sigma}$ field  as follows
\be \label{topdensity}
  q(\mathbf{x}) & = & \displaystyle  \frac{1}{16 \pi^2} \mathrm{tr} \left[ F_{\mu\nu} \tilde{F}^{\mu\nu} \right] 
      =  \displaystyle \frac{-1}{8 \pi^2} \epsilon^{i j k 4} \sum_{a = 1}^{N} F^{(a)}_{j k} F^{(a)}_{i 4} \nonumber \\
    & = & \displaystyle \frac{1 }{8 \pi^2} \sum_{a = 1}^{N} \left< A_{4}^{(a)} \right>
      \left[\epsilon^{ijk} \partial_{i}  F^{(a)}_{j k} (\mathbf{x}) \right]  \nonumber\\
     &=& \left(\frac{g}{2\pi }\right)^2 \cdot\left(\frac{1}{2\pi L}\right)\cdot \sum_{a = 1}^{N} \left< A_{4}^{(a)} \right> \nabla^2  \sigma^{(a)}\nonumber  \\
           &=&\frac{1}{LN} \sum_{a = 1}^{N} \sum_{k = 1}^{M^{(a)}} Q_{k}^{(a)} \delta (\mathbf{r}_{k}^{(a)} - \mathbf{x}).
\ee
The vacuum  expectation value of $  A_{4}^{(a)} $ equals $2\mu^{a}\pi/NL$, where  $\mu^{a}  \alpha_{b} =\delta^a_b$ and it plays the role of the Higgs field in this model 
 as explained in \cite {Shifman:2008ja,Yaffe:2008}.  
One can explicitly see that the   topological charge  in formula (\ref{topdensity})  for a single monopole or antimonopole
is properly normalized\footnote{ We believe a short historical detour on fractionalization of the topological charge in QFT is warranted here. In the given context   fractional topological objects appear in 2 dimensional $CP^{N-1}$ model~\cite{Fateev} which were coined as instanton quarks (instanton partons). These quantum objects carry fractional topological charge $Q=\pm 1/N$, and they are very similar to our monopoles in deformed QCD   discussed in section \ref{model}. These objects do not appear individually in path integral; instead, they appear as configurations consisting $N$ different  $1/N$ objects  such that the total topological charge of each configuration   is integer. Nevertheless, these objects are highly delocalized: they may emerge on opposite sides of the space time or be close to each other with similar   probabilities. 
Similar objects have been discussed in a number of papers in a different context \cite{vanbaal}, \cite{Diakonov:2004jn},
 \cite{Zhitnitsky:2006sr},\cite{Parnachev:2008fy},\cite{Collie:2009iz},\cite{Bolognesi:2011nh}. In particular, it has been argued that the well-established $\theta/N$ dependence in strongly coupled QCD unambiguously implies that the relevant configurations in QCD must carry   fractional topological charges, see review preprint \cite{Zhitnitsky:2006sr} and references therein. The weakly coupled deformed QCD model \cite {Shifman:2008ja,Yaffe:2008} considered in this paper is a precise dynamical realization of this idea.} $Q = \int d^{4}x q(\mathbf{x}) =\pm1/N$. 

Now we introduce truly singlet abelian field $f_{jk} (\mathbf{x}) \equiv \sum_{a = 1}^{N} \left< A_{4}^{(a)} \right>
      \left[    F^{(a)}_{j k} (\mathbf{x}) \right]  $ such that the  topological density operator $q(\mathbf{x})$ for the background monopoles is expressed in terms of this new field as follows, 
\be
\label{f}
q(\mathbf{x}) = \frac{ 1}{16 \pi^2} \mathrm{tr}  F_{\mu\nu} \tilde{F}^{\mu\nu}  
 = \frac{1 }{4 \pi NL}   \left[\epsilon^{ijk} \partial_{i}  f_{j k} (\mathbf{x}) \right]. 
\ee
One should emphasize that the expression on the right hand side  of eq. (\ref{f}) does not represent all the properties of  the topological density operator, e.g. it does not include all non-abelian fluctuations which are present in the system. Rather it should be treated as  a classical background describing the long distance behaviour of $q(\mathbf{x})$. Furthermore, 
  this new abelian field $f_{j k} (\mathbf{x})$ is not an  abelian projection describing the abelian $U(1)^N$ magnetic monopoles in this model. Rather, the combination $\epsilon^{ijk} \partial_{i}  f_{j k} (\mathbf{x})$  describes the gauge invariant  topological density distribution (\ref{f}).
The field  $f_{j k} (\mathbf{x})$ itself is not a gauge invariant object under generic  gauge transformations. Indeed, from relation $q(x) =\partial_{\mu}K^{\mu}(x)$ one can infer that 
  $\epsilon^{ijk}  f_{j k} (\mathbf{x})$ transforms like $K^{i} (x)$. Transformation properties of $K^{\mu}(x)$ are well known;  though this object  
  does not carry the colour indices, it is not a gauge invariant object. In fact, $K^{\mu} (x)$  transforms in a nontrivial way under the large gauge transformations. In particular $\int d^3\sigma_{\mu} K^{\mu} $ determines the winding number of a ``degenerate" vacuum state.  For our specific case of deformed QCD we need to know the  transformation properties for the following operators, 
   \be
\label{f1}
   \left[\epsilon^{ijk} \partial_{i}  f_{j k} (\mathbf{x}) \right] {\rm transforms ~like~ } &~&q( {x})~~~ {\rm (i.e. invariant)}\nonumber \\
    \left[\epsilon^{ijk}    f_{j k} (\mathbf{x}) \right]  {\rm  transforms ~like~ } &~&K^{i} (x).  
   \ee
  Furthermore,  the $f_{jk} (\mathbf{x}) $ field does not discriminate  between different types of monopoles and anti-monopoles  which are classified by the   roots  of the Lie algebra $\alpha_{a} \in \Delta_{\mathrm{aff}}$. Instead, this new field is  sensitive exclusively  to the topological charge density  of these objects  $Q = \int d^{4}x q(\mathbf{x}) =\pm1/N$, not to their abelian magnetic charges $\sim \alpha_{a}  $. As a final comment: there is no   Maxwell dynamical term for this field.    This is  in stark  contrast with  dynamical Maxwell term $  \left(\nabla  \sigma \right)^2$ describing  the abelian magnetic components in the effective action  (\ref{thetaaction}). There is no mystery here as 
  the dynamics of $f_{j k} (\mathbf{x})$ is governed by a pure topological field theory with no Maxwell term. 

One can define a gauge potential $a_i (\mathbf{x})$   associated with the tensor field $f_{j k} (\mathbf{x})$ introduced above 
and a new scalar potential $a(\mathbf{x})$  describing the divergent portion of this tensor as follows:
\be
\label{a}
  f_{j k} (\mathbf{x})  \equiv \left[\partial_{j}a_k(\mathbf{x})-\partial_{k}a_j(\mathbf{x})\right]-\frac{1}{2}\epsilon_{ijk}\partial^{i} a (\mathbf{x}) .   
\ee
The $a_i (\mathbf{x}) $ and  $a(\mathbf{x})$  potentials  in eq.({\ref{a}) are not directly related to the    abelian magnetic potentials
and effective $ \bm{\sigma}$ fields discussed above. 
The fractional topological charge of the monopoles can be expressed in terms of $a(\mathbf{x})$ potential as follows, 
\be
\label{Q}
Q &=&  \int_{\mathbb{R}^{3} \times S^{1}} d^{4}x q(\mathbf{x})    =  \frac{1 }{4 \pi N}  \int_{\mathbb{R}^{3}} d^{3}x  \left[\epsilon^{ijk} \partial_{i}  f_{j k} (\mathbf{x}) \right]\nonumber\\
&=&\frac{-1 }{4 \pi N}  \int_{\mathbb{R}^{3}} d^{3}x \vec{\nabla}^2 a (\mathbf{x})  =\frac{-1}{4 \pi N}\oint_{\Sigma}d\vec{\Sigma}\cdot \vec{\nabla} {a}(\mathbf{x}) , 
\ee
where surface $\Sigma$ defines the boundaries of our system. We have to take this surface to infinity if we define our system on 
$\mathbb{R}^{3}$. Our normalization is chosen in a such a way that a single monopole classified by  $\alpha_a$ and fractional topological charge $Q=  1/N$ is described by an effective long distance field $a (\mathbf{x})$ which satisfies $ \vec{\nabla}^2 a (\mathbf{x}) =-4\pi \delta^3(\mathbf{x})$ with asymptotic behaviour $ a (\mathbf{x})= 1/|\mathbf{x}|$.
 
Our next step is to  insert  the  delta function  into the path  integral with the field $b(\mathbf{x})$ acting as a Lagrange multiplier
\be
\label{delta}
\delta \left( q(\mathbf{x})-  \frac{1 }{4 \pi NL}   \left[\epsilon^{ijk} \partial_{i}  f_{j k} (\mathbf{x}) \right]\right)\sim \nonumber\\
\int {\cal{D}}[b]e^{ i\int d^4x ~b(\mathbf{x})\cdot\left(q(\mathbf{x})-  \frac{1 }{4 \pi NL}   \left[\epsilon^{ijk} \partial_{i}  f_{j k} (\mathbf{x}) \right]\right)}, 
      \ee
     where  $q(\mathbf{x})\sim \mathrm{tr}
		\left[ F_{\mu\nu} \tilde{F}^{\mu\nu} \right] $ in this formula is treated as the original expression (\ref{topologicalcharge}) for the topological density operator 
		including the fast non-abelian gluon degrees of freedom, while $f_{j k} (\mathbf{x}) $ is treated as a slow-varying   external source describing the large distance physics  for a given monopole configuration, similar to the treatment in   \cite{BF} of  external currents for quasiparticles.   
		
		Our task now is to integrate out the original non-abelian fast degrees of freedom and describe the large distance physics in terms of slow varying fields in form of the effective action $S[\bm{\sigma},  f_{j k} , b]$.  We use the same semiclassical approximation 
as   before which is expressed in terms of the   low energy effective action (\ref{thetaincluded}).  The only new element 
in comparison with the previous computations is that  the fast degrees of freedom must be integrated out in the presence of the new slow varying  background fields  $f_{j k} , b$ which appear in  eq. (\ref{delta}). Fortunately, the computations can be easily performed if one notices that the background field $b(\mathbf{x})$ enters eq. (\ref{delta}) exactly in the same manner as $\theta$ parameter enters  (\ref{thetaincluded}).
Therefore, assuming that $b(\mathbf{x})$ is slow varying background field we arrive to the following 
  effective action  
\be
\label{b-action}
&{\cal Z}&\sim\int {\cal{D}}[b]{\cal{D}}[\bm{\sigma}]{\cal{D}}[ f]e^{-S_{\rm top}[b, f]-S_{\rm dual}[\bm{\sigma},   b]}\\
&S&_{\rm top}[b, f]= \frac{i }{4 \pi N}   \int_{\mathbb{R}^{3}}  d^{3}x     b(\mathbf{x})\epsilon^{ijk} \partial_{i}  f_{j k} (\mathbf{x})  \nonumber \\ &=&
 \frac{-i }{4 \pi N}  \int_{\mathbb{R}^{3}}  d^{3}x     b(\mathbf{x})\vec{\nabla}^2 a (\mathbf{x}) ;
\nonumber \\
	&S&_{\rm dual}[\bm{\sigma},   b] = \int_{\mathbb{R}^{3}}  d^{3}x \frac{1}{2 L} \left( \frac{g}{2 \pi} \right)^{2}
		\left( \nabla \bm{\sigma} \right)^{2} \nonumber \\&-& \zeta  \int_{\mathbb{R}^{3}}  d^{3}x \sum_{a = 1}^{N} \cos \left( \alpha_{a} \cdot \bm{\sigma}
		+ \frac{\theta+b(\mathbf{x})}{N} \right).
		 \nonumber
\ee
 There are two new elements in comparison with our previous expression (\ref{thetaaction}). First, the topological term 
 $S_{\rm top}$ emerges. This term can be also written as 
 \be
 \label{bf}
 S_{\rm top}=-i\int_{\mathbb{R}^{3}}   d^{3}x \frac{ b_i(\mathbf{x})\epsilon^{ijk}  f_{j k} (\mathbf{x})}{4 \pi N}, ~b_i(\mathbf{x})\equiv \partial_{i}b(\mathbf{x}), ~~~
 \ee
 which brings\footnote{In fact, the constraints on the field $b_i(\mathbf{x})$ from \cite{BF} require that $b_i(\mathbf{x})\sim \partial_{i}\Lambda$ when the boundary of $\mathbb{R}^{3}$ is a  topologically trivial $S^2$, such that we are not loosing much information with identification 
  $b_i(\mathbf{x})\equiv \partial_{i}b(\mathbf{x})$ in eq. (\ref{bf}).   } it into the line with conventional expression employed  in the Higgs model \cite{BF}.
   The second  new element which appears in (\ref{b-action}) is that   $S_{\rm dual}[\bm{\sigma},   b]$ now depends on pure topological field $b(\mathbf{x})$ which has no Maxwell term. 
 
 One should comment here that we neglected the surface terms in the expression for $S_{\rm top}[b, f]$ such that only the scalar potential 
 $a(\mathbf{x})$ from eq. (\ref{a}) enters the final expression for $S_{\rm top}[b, f]$. These surface terms very often are crucial in similar studies in condensed matter (CM)  systems   defined on a finite manifold with   boundaries  which 
 may have  nontrivial topologies such as a torus. 
 These surface terms are known to be responsible for the dynamics of the so-called ``edge states" in topological field theories. We will return to this point later in   section \ref{analogy} where we discuss the analogy with CM systems. However, for the purpose of this work the surface terms can be neglected as we mostly discuss   a surface with the  trivial topology $\mathbb{S}^2$.  There is no physical  degeneracy in this case  and the system itself  is characterized by a  single unique vacuum state.     Nevertheless, the topological feature of the theory such as the topological long 
 range order described by topological action (\ref{b-action}), (\ref{bf}) persists  when a topologically  trivial manifold is considered. The same system exhibits  some degree of   degeneracy if  $~\mathbb{S}^2\rightarrow \mathbb{T}^2$  as will be discussed in section \ref{analogy}. This degeneracy may bring some new dynamics into the system in terms of the so-called ``edge states" living on the surface.  However, it does not modify  the local correlation functions discussed in this paper. 
 
 The topological features of the system  may have a variety of different manifestations.  In fact, 
  the main goal of the rest of this   section is to argue that the well-known resolution of the celebrated $U(1)_A$ problem is a direct consequence of the topological order described by topological action (\ref{b-action}), (\ref{bf}). In other words, we want to argue that the (would be) Goldstone boson receives its mass  in this system (in apparent    contradiction with conventional  symmetry arguments) as a result of the topological features of the Chern-Simons   action (\ref{bf}).

 \subsection{Topological susceptibility in BF theory.}\label{top_bf}
 We now want to compute the correlation function $\la q(\mathbf{x}), q(\mathbf{0})\ra$   entering the expression for the topological susceptibility (\ref{YM}) by integrating out the $b$ and $a$ fields   
 \be
\la q(\mathbf{x}), q(\mathbf{0})\ra=\frac{1}{{\cal Z}}  \int \frac{{\cal{D}}[b]{\cal{D}}[\bm{\sigma}]{\cal{D}}[ a]e^{-S}\vec{\nabla}^2 a (\mathbf{x}), \vec{\nabla}^2 a (\mathbf{0})}{\left(4 \pi NL\right)^2} .~~~~~~ \nonumber
 \label{top_1}
  \ee
 To carry out  the computations we limit ourself  by considering the $\theta=0$ vacuum state where  $\la\bm{\sigma}\ra=0$ for the  massive $ \bm{\sigma}$ fields. We expand the $\cos$ term in (\ref{b-action})  by keeping the quadratic term for the long range field $b(\mathbf{x})$, 
 \be
  \zeta   \sum_{a = 1}^{N}\cos \left(\frac{b(\mathbf{x})}{N}\right)\simeq \zeta N \left[1- \frac{1}{2}\left(\frac{b(\mathbf{x})}{N}\right)^2\right].
 \ee
 Now, the obtained   Gaussian integral over ${\cal{D}}[b]$ can be explicitly executed, and we are left with the following Gaussian integral over ${\cal{D}}[a]$
  \be
   \label{top_2}
\la q(\mathbf{x}), q(\mathbf{0})\ra&=&\frac{1}{{\cal Z}}  \int {\cal{D}}[ a] \frac{e^{-S[a]}\vec{\nabla}^2 a (\mathbf{x}), \vec{\nabla}^2 a (\mathbf{0}) }{\left(4 \pi NL\right)^2}  \\
S[a]&=& \frac{1}{2N\zeta}\frac{1}{(4\pi)^2}  \int_{\mathbb{R}^{3}}  d^{3}x \left[a (\mathbf{x})\vec{\nabla}^2\vec{\nabla}^2 a (\mathbf{x})\right]  ~~~~\nonumber
  \ee
Next  we rescale the $a (\mathbf{x})$ field
\be
\label{rescale}
a' (\mathbf{x})\equiv \frac{a(\mathbf{x})}{4\pi\sqrt{N\zeta}}
\ee
 to write $S[a]$ in a  more conventional  form
 \be
 \label{a'}
 S[a']= \frac{1}{2 }  \int_{\mathbb{R}^{3}}  d^{3}x \left[a' (\mathbf{x})\vec{\nabla}^2\vec{\nabla}^2 a' (\mathbf{x})\right].
 \ee
With this normalization, the corresponding Gaussian integral over $\int{\cal{D}}[ a']$ can be easily computed 
\be
  \label{top_3}
\frac{\int{\cal{D}}[ a']e^{-S[a']}\left[\vec{\nabla}^2 a' (\mathbf{x}), \vec{\nabla}^2 a' (\mathbf{0})\right]}{\int{\cal{D}}[ a']e^{-S[a']}}=\delta^3(\mathbf{x}), 
\ee
where $S[a']$ is defined by eq. (\ref{a'}). 
 Now we are ready to complete the    computations of  the topological susceptibility   using  the topological BF action  (\ref{b-action}).  
We  express the original topological density operator (\ref{Q}) in terms of $a(\mathbf{x})$ and take into account the expression for the Gaussian integral (\ref{top_3}). The final expression  for the gauge invariant correlation function 
\be
\label{YM_top}
\la q(\mathbf{x}), q(\mathbf{0})\ra =\frac{\zeta}{NL^2}\delta^3(\mathbf{x})
\ee
  precisely reproduces the original formula  (\ref{YM}) which was   derived   without   mentioning of any auxiliary fields~\cite{Thomas:2011ee}. 
 One can also compute a gauge variant  correlation function  
 \be
 \label{K_top}
 \lim_{k\rightarrow 0} \int d^3x e^{ikx} \la \nabla_i a(\mathbf{x}), \nabla_j a(\mathbf{0})\ra \sim\frac{k_i k_j}{k^4}.
 \ee
 This object is very similar   to the computations (\ref{K}) using the Veneziano ghost.   The unphysical pole (\ref{K_top}) has precisely the same nature as the pole  in the Veneziano construction  (\ref{K}). In fact, the transformation 
 properties of 
 our field $ \nabla_i a(\mathbf{x}) $ are the same as the $K_{i} (\mathbf{x})$ field (\ref{f1}) in the Veneziano construction (\ref{K}).
 
  Based on this observation and  comparing (\ref{K}) with (\ref{K_top}) we identify our topological fields constructed for the deformed QCD   with the effective Veneziano ghost.  This identification uncovers the nature of the Veneziano ghost as an effective topological non-propagating field.  
 In both cases this pole is not related to any physical massless  degrees of freedom which are not present in the system as the theory is gapped. Rather, it contributes to the  non-dispersive portion of the gauge invariant correlation function (\ref{top1}), (\ref{K1}), (\ref{YM_top}).
   Still, this unphysical topological field does contribute to the $\theta$ dependent portion of the ground state energy.

  In weakly coupled deformed QCD one can  carry out all the computations without even mentioning the topological fields or the Veneziano ghost as formula (\ref{YM}) shows. However, the formulation of this phenomenon in terms of the topological QFT reveals its deep nature which is otherwise hard to understand. As explained above this contact term (\ref{YM_top}) is not related to any physical propagating degrees of freedom.  In computations (\ref{YM})  it emerges as a result of the tunnelling    transitions between the degenerate topological sectors.  The  non-dispersive nature of this term  in the present computations (based on the topological effective Lagrangian (\ref{b-action}))    manifests  itself  in the saturation of   $\chi_{YM}$   thorough the non-propagating, non-dynamical  long range $b(\mathbf{x})$ and  $a(\mathbf{x})$ fields. These fields are not dynamical fields as they do not have the Maxwell term.
   Nevertheless, these fields  are crucial as they saturate the non-dispersive contact term in the topological susceptibility.

Entire  framework   advocated in this paper is in fact a matter of convenience rather than necessity. The same comment  also applies  to CM systems: the BF formulation~\cite{Wen:1989iv,Wen:1990zza,Moore:1991ks,BF,Cho:2010rk,Wen:2012hm} using the topological QFT is simply a matter of convenience  to represent the known and previously established results (  braiding phases, the ground state degeneracy, etc).      As we shall discuss in section \ref{analogy} the manifestations of this long range order  in QCD and in CM systems are somewhat different, but the beauty of the topological BF formulation remains the same.

  \subsection{The  mass generation for  the (would be) Goldstone boson in a topologically ordered system}\label{eta}

We want to reproduce the behaviour (\ref{chi_QCD}) in deformed QCD using an appropriate generalization of the topological BF action (\ref{b-action}) when the massless quarks are introduced into the system. This study will further  illuminate the relation between  the auxiliary $a(\mathbf{x}), b(\mathbf{x})$ fields and the unphysical Veneziano ghost.  To proceed with  this task  we have to introduce the light matter field which is represented in this model by  the $\eta'$ -field \cite{Thomas:2011ee}.  If $\chi_{YM}$ were to vanish the $\eta'$ would be the conventional  massless Goldstone boson which is nothing but the phase of the chiral condensate. However, $\chi_{YM}\neq 0$  in  4d QCD (\ref{top1})  as well as  in deformed QCD  (\ref{YM}).
 In other words, 
the $\eta'$ field receives its mass exclusively  as a result of generating of the nonzero contribution to $\chi_{YM}$ with the ``wrong sign" (\ref{top1}),  which is the key element in the  resolution of the celebrated  $U(1)_A$ problem~\cite{witten,ven,vendiv,Rosenzweig:1979ay,Nath:1979ik,Kawarabayashi:1980dp}.  In the context of the present work we want to see how the $\eta'$ physical contribution exactly cancels 
(as the Ward Identities require) the non-dispersive term in  
the topological density   (\ref{chi_QCD})  in the BF formulation using the topological action (\ref{b-action}). It will shed a new    light on a very deep  relation, already mentioned above,   between the Veneziano ghost and the topological $a(\mathbf{x}), b(\mathbf{x})$ fields.  Essentially we want to see how the (would be) Goldstone boson becomes a massive field in topological QFT in apparent contradiction with conventional  symmetry arguments. 

 In the dual sine-Gordon theory   the $\eta'$ meson field   appears  exclusively in combination with the $\theta$ parameter as $\theta \rightarrow \theta - \phi (x)$, where $\phi$ is the phase of the chiral  condensate which, up to dimensional normalization parameter,  is identified with physical  $\eta'$ meson in QCD. As it is well known, this property is the direct result of the transformation properties of the path integral measure under the chiral transformations $\psi\rightarrow \exp(i \gamma_5\frac{\phi}{2})\psi$.  Therefore, $\phi (x) $ enters the effective action exactly in the same way as the $ b(\mathbf{x})$ field does   (\ref{delta}).  Therefore,  
we can  integrate out the fast degrees of freedom exactly   the   way   we did previously  to arrive at  the following effective action
which now includes the (would be) Goldstone boson field
\be
\label{eta-action}
&{\cal Z}&\sim\int {\cal{D}}[b]{\cal{D}}[\bm{\sigma}]{\cal{D}}[ a]{\cal{D}}[\phi]e^{-\left(S_{\rm top}+S_{\rm dual}[\bm{\sigma}, b, \phi]+S_{\phi}\right)}\\
&S&_{\phi}= \int_{\mathbb{R}^{3}}  d^{3}x \frac{c}{2}\left( \nabla \phi \right)^{2}\nonumber\\
&S&_{\rm top}[b, a]=
 \frac{-i }{4 \pi N}  \int_{\mathbb{R}^{3}}  d^{3}x     b(\mathbf{x})\vec{\nabla}^2 a (\mathbf{x}) ;
\nonumber \\
	&S&_{\rm dual}[\bm{\sigma},   b, \phi] = \int_{\mathbb{R}^{3}}  d^{3}x \frac{1}{2 L} \left( \frac{g}{2 \pi} \right)^{2}
		\left( \nabla \bm{\sigma} \right)^{2} \nonumber \\&-& \zeta  \int_{\mathbb{R}^{3}}  d^{3}x \sum_{a = 1}^{N} \cos \left( \alpha_{a} \cdot \bm{\sigma}
		+ \frac{\theta+b(\mathbf{x})-\phi(\mathbf{x}) }{N} \right),
		 \nonumber
\ee
where coefficient $c$ determines the normalization of the $\phi$ field, and $c/L $ plays the role of $f_{\eta'}^2$ in conventional QCD,  while the $\eta'$ mass is expressed in terms of this coefficient  by eq. (\ref{mass}).
 
There are two new elements which appear in (\ref{eta-action}) in comparison with (\ref{b-action}). First, kinetic term for the $\phi$ field is generated and parametrized by $S_{\phi}$. Secondly,  the  $\phi(\mathbf{x}) $ field enters the $S_{\rm dual}[\bm{\sigma},   b, \phi]$ in a   specific way consistent with the Ward Identities.

Our task now is to compute the topological susceptibility in the presence of the massless quark field by integrating  out the auxiliary $b(\mathbf{x}), a(\mathbf{x})$ fields. We repeat the same steps which led us to (\ref{top_2}), with the  only difference being  that the $a(\mathbf{x})$ and $\phi (\mathbf{x})$ fields    mixing such that effective action $S_{QCD}[a, \phi]$ now takes the form
 \be
 \label{top_QCD}
&\,&\la q(\mathbf{x}), q(\mathbf{0})\ra_{QCD}=\frac{1}{{\cal Z}}  \int \frac{{\cal{D}}[ a]e^{-S_{QCD}}\left[\vec{\nabla}^2 a (\mathbf{x}), \vec{\nabla}^2 a (\mathbf{0})\right] }{\left(4 \pi NL\right)^2} \nonumber\\
&\,&S_{QCD}[a, \phi]=   \frac{1}{2N\zeta (4\pi)^2} \int_{\mathbb{R}^{3}}  d^{3}x  \left[a (\mathbf{x})\vec{\nabla}^2\vec{\nabla}^2 a (\mathbf{x})\right] \nonumber\\
&\,&+ \int_{\mathbb{R}^{3}}  d^{3}x \left[\frac{c}{2}\left( \vec{\nabla} \phi(\mathbf{x}) \right)^{2}+\frac{i}{4\pi N} \vec{\nabla}\phi (\mathbf{x})\cdot \vec{\nabla} a (\mathbf{x})\right]   ,
  \ee
where we performed   the  integration  by parts  on  the mixing term $\int   d^{3}x \phi  \vec{\nabla}^2 a =- \int  d^{3}x \vec{\nabla}\phi \cdot \vec{\nabla} a$. Our next step is to eliminate the non-diagonal  term $ \int  d^{3}x \vec{\nabla}\phi \cdot \vec{\nabla} a$ in (\ref{top_QCD}) by making a shift 
\be
\label{phi_2}
\frac{\phi_2 (\mathbf{x})}{\sqrt{c}} \equiv \phi (\mathbf{x})+\frac{i}{4\pi c N}a (\mathbf{x}),
\ee
where ${1}/{\sqrt{c}}$ is inserted to the left hand side of eq. (\ref{phi_2}) in order to bring the kinetic term for $\phi_2 (\mathbf{x})$ field to its canonical form.
The problem  is reduced to the computations of the Gaussian integral similar to  (\ref{top_2}) with the only difference being that the effective action now, after the rescaling (\ref{rescale}),  takes the form
  \be
   \label{S_QCD}
S_{QCD}[a' , \phi_2] &=&  \frac{1}{2} \int_{\mathbb{R}^{3}}  d^{3}x\left( \vec{\nabla} \phi_2(\mathbf{x}) \right)^{2}\\
&+&\frac{1}{2 }  \int_{\mathbb{R}^{3}}  d^{3}x a' (\mathbf{x})\left[ \vec{\nabla}^2\vec{\nabla}^2- m_{\eta'}^2 \vec{\nabla}^2 \right]a' (\mathbf{x})\nonumber
    \ee
 which replaces previous expression (\ref{a'}). In formula (\ref{S_QCD}) parameter $m_{\eta'}^2$ is the $\eta'$ mass in this model and it is  given by eq.(\ref{mass}). In terms of this rescaled field  $a' (\mathbf{x})$   the  Gaussian integral which enters (\ref{top_QCD}) can be easily computed and it is given by
\be
\label{top_4}
  \frac{\int{\cal{D}}[ a']e^{-S_{QCD}}\vec{\nabla}^2 a' (\mathbf{x}), \vec{\nabla}^2 a' (\mathbf{0})}{\int{\cal{D}}[ a']e^{-S_{QCD}[a']}}= \left[ \delta(\bold{x})-m_{\eta'}^2G_{m_{\eta'}}(\bold{x}) \right]  ~~~~~
\ee
where $S_{QCD}$ is defined by (\ref{S_QCD}) and  the massive Green's function
 $G_{m_{\eta'}}(\bold{x})=\frac{e^{-m_{\eta'}r}}{4\pi r} $ is normalized in conventional way ($m_{\eta'}^2\int d^3xG_{m_{\eta'}}(\bold{x})=1$). Collecting all numerical coefficients from (\ref{rescale}), (\ref{top_QCD})  and (\ref{top_4}) the final expression for the topological susceptibility in the presence of massless quark    takes the form 
 \be
\label{QCD_final}
\la q(\mathbf{x}), q(\mathbf{0})\ra_{QCD} =\frac{\zeta}{NL^2}\left[ \delta(\bold{x})-m_{\eta'}^2\frac{e^{-m_{\eta'}r}}{4\pi r} \right]. 
\ee
This precisely reproduces our previous formula (\ref{chi_QCD}) which was derived by explicit integration over all possible monopole's configurations
without even mentioning the topological auxiliary fields. It now has two terms: the first term with the ``wrong sign" which has non-dispersive nature, and which was computed previously (\ref{YM_top}) using the same auxiliary topological fields. It also has a new   dispersive term related to the massive $\eta'$ propagating degrees of freedom. It has a negative sign in accordance with general principles (\ref{G}). The Ward Identities are satisfied $\chi_{QCD}=
\int d^3x \la q(\mathbf{x}), q(\mathbf{0})\ra_{QCD}=0$ as a result of exact cancellation between the two terms. The celebrated  $U(1)_A$ problem
is resolved in this framework exclusively as a result of dynamics of the topological  $a(\mathbf{x}), b(\mathbf{x})$ fields. These fields are not propagating degrees of freedom, but nevertheless  
generate  a crucial non-dispersive contribution with the ``wrong sign" which is the key element for the formulation and resolution of the $U(1)_A$ problem and the generation of the $\eta'$ mass.

\subsection{ Topological fields and  the Veneziano ghost.}\label{KS_ghost}  
The expression for the correlation  function (\ref{top_4}) with action (\ref{S_QCD}) can be represented in a complementary  way which makes the
connection between the Veneziano ghost and topological fields much more explicit. 
 
 To proceed with out task,  we use a standard  trick to represent the 4-th order operator $\left[ \vec{\nabla}^2\vec{\nabla}^2- m_{\eta'}^2 \vec{\nabla}^2 \right]$ which enters the effective action (\ref{S_QCD}) 
    as a combination of  two terms with the opposite signs: a ghost field $\phi_1$ 
and a massive physical $\hat\phi $ field. To be more specific, we write 
\be
\label{inverse}
\frac{1}{\left[ \vec{\nabla}^2\vec{\nabla}^2- m_{\eta'}^2 \vec{\nabla}^2 \right]}= \frac{1}{m_{\eta'}^2}\left(\frac{1}{ \vec{\nabla}^2 - m_{\eta'}^2 } -  \frac{1}{ \vec{\nabla}^2 } \right) , 
\ee
such that the Green's function for the $a(\mathbf{x})$ field which enters the expression for the topological susceptibility (\ref{top_4}) can be represented as a combination of two Green's functions,  for the physical massive field with conventional kinetic term and for the ghost field with the ``wrong" sign for the kinetic term. 
  As usual, the presence of 4-th order operator in eq. (\ref{S_QCD})  is a signal that the ghost is present in the system. This signal   is explicit  in eq. (\ref{inverse}).    The   contact term     in this framework is  represented 
by the ghost contribution.  
 Indeed,   the   relevant correlation function   which enters the expression for the topological susceptibility (\ref{top_4}) can be explicitly computed using expression (\ref{inverse}) for the  inverse operator  as follows 
 \be\label{QCD_ghost}
&\,& \frac{\int{\cal{D}}[ a']e^{-S_{QCD}[a']}\vec{\nabla}^2 a' (\mathbf{x}), \vec{\nabla}^2 a' (\mathbf{0})}{\int{\cal{D}}[ a']e^{-S_{QCD}[a']}} \\
&=& \int \frac{\dd^3p  }{\left(2\pi\right)^3} e^{-i p x} \frac{p^4}{m_{\eta'}^2}\left[ - \frac{1}{p^2+m_{\eta'}^2} + \frac{1}{p^2} \right] 
\nonumber \\
   &=&\int \frac{\dd^3p}{\left(2\pi\right)^3}  e^{-i p x} \left[ \frac{p^2}{p^2+m_{\eta'}^2}   \right] = \left[ \delta(\bold{x})-m_{\eta'}^2\frac{e^{-m_{\eta'}r}}{4\pi r} \right],  \nonumber
  \ee
 which, of course, is the same final expression  we had before (\ref{top_4}) with the only difference being that it is now explicitly expressed  as a combination of two terms: a physical massive $\eta'$ contribution and an unphysical contribution which saturates the contact term with the ``wrong" sign.
 
 Now we  represent the  correlation function  (\ref{QCD_ghost}) by introducing two fields $\phi_1(\mathbf{x})$ and $\hat{\phi}(\mathbf{x})$
 replacing the $a'(\mathbf{x})$ which enters the effective action (\ref{S_QCD}) as the 4-th order operator.  To be more precise, we rewrite our action (\ref{S_QCD}) in terms of these new fields  $\phi_1(\mathbf{x})$ and $\hat{\phi}(\mathbf{x})$
 as follows
 \be
   \label{S_ghost}
S_{QCD}[\hat{\phi}, \phi_1,\phi_2] =  \frac{1}{2} \int_{\mathbb{R}^{3}}  d^{3}x\left[\left( \vec{\nabla} \phi_2(\mathbf{x}) \right)^{2}-\left(\vec{\nabla} \phi_1(\mathbf{x}) \right)^{2} \right]\nonumber\\
+\frac{1}{2 }  \int_{\mathbb{R}^{3}}  d^{3}x \left[ \left( \vec{\nabla} \hat{\phi}(\mathbf{x}) \right)^{2}+ m_{\eta'}^2  \hat{\phi}^2(\mathbf{x})\right] 
   ~~~~~ \ee
with the  $a'(\mathbf{x})$ field   expressed in terms of the new fields  $\phi_1(\mathbf{x})$ and $\hat{\phi}(\mathbf{x})$ as  
\be
\label{a-ghost}
a'(\mathbf{x})\equiv \frac{1}{m_{\eta'}}\left(\hat{\phi}(\mathbf{x})-\phi_1(\mathbf{x})\right), 
\ee
 while the topological density $q(\mathbf{x})$  operator is expressed in terms of these fields as  
 \be
 \label{q_ghost}
 q =\sqrt{\frac{\zeta}{NL^2}} \vec{\nabla}^2a' =\sqrt{\frac{\zeta}{NL^2m_{\eta'}^2}} \vec{\nabla}^2\left(\hat{\phi}-\phi_1 \right).
 \ee

 This redefinition obviously leads to our previous result (\ref{QCD_final}), (\ref{QCD_ghost}) when we use the Green's functions determined by the Lagrangian (\ref{S_ghost})  for the physical massive field $\hat{\phi}$ and the ghost $\phi_1$, 
  \be
\label{QCD_final_g}
\la q(\mathbf{x}), q(\mathbf{0})\ra_{QCD} =\frac{\zeta}{NL^2}\left[ \delta(\bold{x})-m_{\eta'}^2\frac{e^{-m_{\eta'}r}}{4\pi r} \right]. 
\ee
 It is quite amazing that precisely this structure (\ref{S_ghost}) had emerged previously in the study of of the $U(1)_A$ problem  in the  Kogut-Susskind (KS) formulation of the 2d Schwinger model~\cite{KS}, see also \cite{Zhitnitsky:2011tr} with related discussions. The topological density operator in the 2d Schwinger model  $\epsilon_{\mu\nu} F^{\mu\nu} $ is also expressed as 
 $\epsilon_{\mu\nu} F^{\mu\nu} \sim \partial_{\mu}\partial^{\mu} \left(\hat{\phi}-\phi_1 \right)$ similar to (\ref{q_ghost}).
 Furthermore, one can show that  this structure still holds even with non-vanishing  quark mass $m_q$, in which case an additional  term $\sim m_q\cos(\hat{\phi}+\phi_2-\phi_1)$ appears
 in effective action (\ref{S_ghost}), similar to  analogous expression in the KS description~\cite{KS}. 
 In fact, our notations for the $\hat{\phi}, \phi_1,\phi_2$ fields entering (\ref{S_ghost}) are precisely the same as those    used in ref. \cite{KS} to emphasize the similarity. Furthermore, an analogous   structure also emerges in 4d QCD when the the topological density operator is expressed in terms of the Veneziano ghost   where $q\sim \Box  (\hat{\phi}-\phi_1)$ has precisely the same structure \cite{Zhitnitsky:2010zx}.
  
 An important  point here is that the contact term in this framework is explicitly saturated by the topological non-propagating auxiliary fields expressed in terms of the ghost field $\phi_1$, similar to the 2d KS ghost  or the 4d Veneziano ghost 
 \footnote{\label{ghost}It is important  to emphasize that KS and Veneziano  ghosts should  not be confused with 
the conventional Fadeev-Popov ghost which is normally introduced into the theory
to cancel out unphysical polarizations of the gauge fields. Instead, the KS, Veneziano  ghost is introduced to account for the  existence of  topological sectors
in the theory, see \cite{Zhitnitsky:2011tr} for references and details. In the four dimensional case the Veneziano ghost can not be confused with the Fadeev-Popov ghost as the Veneziano ghost being a singlet   does not carry  a colour index, in contrast  with  the Fadeev-Popov ghost.
 The sole purpose of the Veneziano ghost is to saturate 
 the contact term with the ``wrong sign" in the topological susceptibility, similar to eq. (\ref{QCD_final_g}) in the deformed QCD model.}. From our original formulation without  any 
auxiliary fields reviewed in section \ref{top} it is quite obvious that our theory is unitary and causal. When we introduce  the auxiliary fields (which are extremely useful when one attempts  to study the long range order) the unitarity, of course, still holds.    Formally, 
the   unitary holds  in this formulation because  the ghost field $\phi_1$ is always paired up with $\phi_2$ in   every gauge invariant matrix element as explained in ~\cite{KS} (with the only exception being the topological density operator (\ref{q_ghost}) which requires a special treatment presented in  this section).  The condition that enforces this statement is the Gupta-Bleuler-like condition on the physical Hilbert space ${\cal H}_{\mathrm{phys}}$ which reads  
\be\label{gb}
(\phi_2 - \phi_1)^{(+)} \left|{\cal H}_{\mathrm{phys}}\right> = 0 \, ,
\ee
where the $(+)$ stands for the positive frequency Fourier components of the quantized fields.  
The crucial point here is that the formulation of the theory using the topological fields has an enormous advantage as the long range order is explicitly present in the formulation (\ref{b-action}), (\ref{eta-action}) and therefore, in  the equivalent formulation   in terms of the ghost field as eq. (\ref{S_ghost}) states.  

$\bullet$ Our arguments, based on  analysis of   a simplified version of QCD essentially suggest  that the key element of the $U(1)_A$ problem represented by eq.(\ref{YM})   is a direct manifestation  of the long range  topological action (\ref{b-action}), (\ref{bf}), (\ref{eta-action}). A similar structure in CM systems is known 
to describe  topologically ordered phases. It is natural to assume that the deformed QCD also belongs to a topologically ordered phase. Furthermore, one can explicitly see from our computations above that  the $\eta'$ generates its mass as a result of a mixture  of  the (would be) Goldstone field with the topological auxiliary field. 
 Therefore, we interpret the well known resolution of the $U(1)_A$ problem in deformed QCD as  a {\it result of dynamics of  a topological field} describing the long range order of the  system.

\section{ Physical Interpretation. Analogies with condensed matter systems.}\label{analogy}
The central line  of all our previous discussions is that the resolution of the celebrated $U(1)_A$ problem in deformed QCD is a direct manifestation of the topological order of the system. The main argument is based on the analysis of the topological action  (\ref{b-action}), (\ref{bf}), (\ref{eta-action})  which exactly reproduces
the crucial element of the $U(1)_A$ problem, the topological susceptibility as eqs. (\ref{YM_top}), (\ref{QCD_final}), (\ref{QCD_final_g}) demonstrate. A similar topological action in condensed  matter (CM) systems  is known to lead to a variety of very non-trivial properties  as a manifestation of  topologically ordered phases realized in these systems, see 
\cite{Wen:1989iv,Wen:1990zza,Moore:1991ks,BF,Cho:2010rk,Wen:2012hm} and many references therein. 
 
 \subsection{Differences and Similarities between CM systems and deformed QCD}

 As we already mentioned,  there is a fundamental difference between   CM systems defined in Minkowski space time  $d=(2+1)$ and Euclidean 3d 
``deformed QCD"  which has been studied in the present work. 
In particular, instead of propagating quasiparticles in CM systems we have pseudo-particles (monopoles) which saturate the path integral. As a result of this difference we can not use many  standard tools which normally would  detect the topological order. For example, we can not compute 
 the braiding phases of charges and vortices 
which are normally used in CM systems simply because our system does not support    such kind of excitations. Further to this point, the  topological $b(\mathbf{x}), a(\mathbf{x})$ fields  entering the abelian BF  action 
(\ref{b-action}), (\ref{bf}), (\ref{eta-action})  are not related in any way to the physical $E\&M$ field in contrast with CM systems where the topological $a_{\mu}$ field couples to  the physical $E\&M$ charges. This prevents us from forming a real vortex  which carries  magnetic field. Also, these  topological $b(\mathbf{x}), a(\mathbf{x})$ fields do not have canonical Maxwell terms. Instead, the abelian topological field  in our case decouples from $E\&M$ charges, and behaves like  the $K_{\mu}$ field as discussed in section \ref{BF}, see eq.(\ref{f1}).  

Furthermore, the ``degeneracy"  in the deformed QCD model
  is related to the degeneracy of winding states $| n\ra$ which are connected to each other by large gauge transformation, and therefore
 must be identified as they  correspond to   the same physical state.
     It is very different from  the conventional term ``degeneracy" in  topologically ordered CM systems  wherein {\it distinct} degenerate states are present in the system as a result of formulation of a theory on a topologically non-trivial manifold such as a torus. 
     
     In the case of deformed QCD one should anticipate a similar behaviour when the system is defined on topologically nontrivial manifold. To be more specific,  if 
   the boundary of   the Euclidean space       $\mathbb{R}^{3}$ in eq.(\ref{standardYM}) is a topologically trivial $\mathbb{S}^{2}$ than one should expect a unique vacuum state. 
    At the same time, if  the Euclidean space       $\mathbb{R}^{3}$ in eq.(\ref{standardYM})   is additionally compactified  on a large torus, i.e.
     $\mathbb{R}^{3}\rightarrow \mathbb{T}^{2}\times \mathbb{R}^{1}$, than one should expect an additional topological $\mathbb{Z}_{2}$ symmetry for $SU(2)$,  in close analogy with the behaviour of a CM system~\cite{BF}.  In fact, the emergence of this additional topological $\mathbb{Z}_{2}$ symmetry  has been explicitly demonstrated  in deformed QCD in~\cite{Anber:2011gn}. This symmetry is realized as  a {\it physical} degeneracy in the limit of large $\mathbb{T}^{2}$.  
  The    additional topological $\mathbb{Z}_{2}$ symmetry emerges in the system defined on $\mathbb{T}^{2}$
  as a result of nontrivial holonomy (nontrivial Polyakov loop) characterized by the system. For $\mathbb{S}^{2}$ the holonomy is obviously trivial, therefore the vacuum state is unique.  
     
     Such a behaviour of the system further  supports our claim  that the deformed QCD  belongs to a topologically ordered phase. Indeed, a topological phase in a gapped CM system is normally characterized by a set of degenerate ``pseudo-ground" states 
     with the energy difference $\delta E\sim \exp(-L)$, where $L$ is the size of the system \cite{Wen:1989iv,Wen:1990zza,Moore:1991ks,BF,Cho:2010rk,Wen:2012hm}. The deformed QCD defined on a  large torus of size $L$ obviously satisfies this property~\cite{Anber:2011gn}. Furthermore, the vacuum expectation values for all {\it local} operators for these degenerate sates in CM systems are the same (up to exponentially small corrections  $\exp(-L)$). This property is also satisfied  in the deformed QCD defined on the torus. These  characteristics    are considered as the unambiguous signatures of topological order. The corresponding features are in drastic contrast   with a conventional gapped theory which normally is not  sensitive    to any variations of the boundary conditions at arbitrary large distances.

    The moral is that   the systems in both cases (deformed QCD vs CM) demonstrate a huge sensitivity to the large distance physics formulated in terms of the boundary conditions. When  we define a  system on a topologically trivial manifold such as sphere, there would be a unique  ground state   in this ``trivial" case.  However, the  long range order  in this   system   governed by  topological BF action persists. One can not use the conventional tools, such as degeneracy of the ground state, to detect a topological order. Instead, one should use some  different observables to detect the long range order which obviously remains in the system even if it is formulated on a topologically trivial manifold.   As we advocated above, the resolution of the $U(1)_A$ problem formulated in terms of the long range topological fields is another manifestation of the topological order.

    \subsection{``String-net condensation" in CM \cite{Levin:2004mi} vs  long range   ``Skeleton" in the lattice QCD \cite{Horvath:2003yj,Horvath:2005rv}}\label{skeleton}
 While we demonstrated a number of supporting arguments  that   QCD indeed lies in a topologically ordered phase, its  microscopic  nature remains a mystery.  On other hand, in CM systems   it has been suggested  \cite{Levin:2004mi}  that the microscopic  mechanism for topological phases is the so-called ``String-net condensation". 
 In this section we speculate that the structure which has  been observed in the lattice QCD \cite{Horvath:2003yj,Horvath:2005rv}} plays the same role as ``String-net condensation" in CM, and therefore, it provides a
 microscopic 
 mechanism leading to the long range topological order in QCD.

Indeed, in both cases the configurations  themselves  have lower dimensionality than the space itself. However, these low-dimensional  configurations are so dense, and they fluctuate so strongly that they almost fill the entire space. 
In both cases an effective  tension of the configurations vanishes as a result of large entropies of the objects which overcome the internal tension. This leads to the condensation 
of the ``string-nets" and percolation of the ``Skeleton" correspondingly. If the effective tensions of these configurations did not vanish, we would observe a finite number of fluctuating objects with finite size in the system instead of observed percolation of the ``Skeleton" and condensation of the ``string-nets".
Furthermore,  typically  a ``Skeleton"   spreads  over maximal   distances percolating through the entire volume of the system similar to ``string-nets"  which condense.  Finally, in both cases the ${\cal P}$ and ${\cal T}$ invariance holds as a result of presence of two coherent  networks. In  String -net condensation this is achieved by considering two topological QFT's with opposite chiralities.
 In ``Skeleton" studies  there are two oppositely- charged sign-coherent connected structures (ÒsheetsÓ).  The ${\cal P}$ and ${\cal T}$ invariance holds in QCD as a result of  delicate cancellation between the opposite sign interleaved sheets. This delicate cancellation may be locally violated as a result of an external impact such as heavy ion collision.  Apparently, such a   local ${\cal P}$ and ${\cal CP}$ violation has indeed   been observed in heavy ion collision experiments at RHIC, Brookhaven, and LHC, Geneva, see some  comments in  conclusion.

The crucial difference between the two cases is of course the nature of the objects: CM systems live in  real Minkowski space-time while lattice QCD measurements are done in Euclidean space-time where the corresponding configurations    saturate the path integral. In the present context it implies that while in CM systems the corresponding ``string -nets" are made of real particles/quasiparticles 
organized into extended objects which may condense,   in QCD the corresponding extended ``skeleton"  configurations  live in 4d Euclidean space. Therefore, they should be interpreted as the objects describing the    tunnelling events in Minkowski space time, similar to instantons. The term ``condensation" normally used in CM literature is not quite appropriate for such 4d objects. Therefore, it is more appropriate to describe  the relevant physics by the term  ``percolation" of extended configurations. It is clear that much work needs do be done before this speculation becomes a workable framework.

    \section{Conclusion. Speculations. Future Directions}\label{conclusion}
  The main ``technical" result of the  present work can be formulated as follows. The  analysis of   a simplified version of QCD   suggests  that   the non-dispersive contribution to the topological susceptibility  (\ref{top1})
  (which is  a key element for the   formulation and resolution of celebrated  $U(1)_A$ problem)  emerges because   the deformed QCD can be     described in terms of  the auxiliary topological $a(\mathbf{x}), b(\mathbf{x}) $ fields governed by the BF-like action (\ref{b-action}), (\ref{bf}), (\ref{eta-action}).  Such an effective action in CM systems is normally considered  a signal for the emergence of a topologically ordered phase with a number of known striking features. Therefore, it is naturally to assume that the deformed QCD also lies in a topologically ordered phase.  While  we can not use many standard tools (such as  the braiding phases, physical degeneracy etc as explained in section \ref{analogy}) which would  confirm  the topological order,     we still can see the long range order in the system through different phenomena  discussed in this paper. One could argue that this long range order persists in strongly coupled QCD as well, as there should not be  any phase transitions in the passage  from  ``deformed" to real QCD.

 However, the computations in ``deformed QCD" do not say much about the source, 
 the nature, the ``mechanism"   of this long range order.   At this point 
  we  return to our first line in the introductory  section \ref{introduction} on  puzzling  recent lattice results \cite{Horvath:2003yj,Horvath:2005rv,Horvath:2005cv,Alexandru:2005bn,Ilgenfritz:2007xu,Ilgenfritz:2008ia,Bruckmann:2011ve,Kovalenko:2004xm,Buividovich:2011cv} which   actually motivated this study. 
    In CM systems it has been shown that a physical mechanism for topological phases can be formulated in terms  of the string-net condensation \cite{Levin:2004mi}.  We speculated  in section \ref{skeleton} that the structure observed on the  lattices and coined  ``skeleton" is analogous to ``String-net condensation" in CM systems. In other words, we speculate   that the microscopic mechanism for the long range order in QCD  could be precisely the long range structure which has   already   been observed   \cite{Horvath:2003yj,Horvath:2005rv,Horvath:2005cv,Alexandru:2005bn,Ilgenfritz:2007xu,Ilgenfritz:2008ia,Bruckmann:2011ve,Kovalenko:2004xm,Buividovich:2011cv}. 
  
  However, the observable  manifestations of this topological ordered phase are much more difficult to detect as explained in \ref{skeleton} because an $E\&M $  does not couple directly to the topological structure  in contrast with CM systems where an external $E\&M$ field is a perfect probe of 
  the long range order. Nevertheless, there could be  some other  manifestations of the long range order such as generating the mass of  would be Goldstone boson (the so-called $U(1)_A$ problem in QCD considered in the present work) or others which have not been discovered yet. 
  We would like to present a (very speculative) list of possible manifestations of the QCD long range order, if it is   confirmed by future analytical and numerical studies:
  
   1)   Over the years, in hadron production studies in a variety of high energy collision experiments  from $e^+e^-$  annihilation to $pp$ and $p\bar{p}$  interactions with energies from a few GeV up to the TeV range, the production pattern always shows striking thermal aspects with an  apparently   universal temperature around  $T_H\simeq (160- 170) ~{\text MeV}$.   It is very difficult to understand the  nature of this ``apparent thermalization" as  one can not even speak   about kinetic equilibration.    In other words, the thermal spectrum in all  high energy collisions emerges in spite of the fact that the statistical thermalization could never be reached in those systems. Hagedorn concluded that the hadrons are simply Òborn in equilibriumÓ \cite{Hagedorn:1965st}, in apparent contradiction with causality.  
   
   We would like to speculate  that the source for  such  striking thermal features  could be related to the coherent structure of sheets making the ``skeleton". In this case the 
``apparent thermalization" would emerge  as a result of tunnelling events  accompanied by particle production, rather than a result of  interaction of the produced particles, see \cite{Castorina:2007eb,Zhitnitsky:2010zx,Zhitnitsky:2012im} and references therein where the very first steps along this direction 
  have been undertaken.  We believe that the crucial element of this idea (which is formulated in terms of the tunnelling events  of the {\it long range} coherent ``skeleton" described by the auxiliary axion  field,  similar to $a(\mathbf{x}), b(\mathbf{x})$   fields from section \ref{BF-section})  has the potential  to eventually explain  Hagedorn's notion that the states were prepared before the collision occurs. This explanation would not   violate causality as topological auxiliary   fields are not propagating degrees of freedom in this framework.

2) The violation of local ${\cal P}$ and ${\cal CP}$ invariance in QCD  has been a subject of intense studies for the last couple of years as a result of very  interesting ongoing   experiments at  RHIC (Relativistic Heavy Ion Collider)~\cite{Abelev:2009tx,Wang:2012qs} and, more recently, at  the LHC (Large Hadron Collider)~\cite{Selyuzhenkov:2012py,Abelev:2012pa,Voloshin:2012fv,Selyuzhenkov:2012mf}. The main idea to explain the observed asymmetries is to 
       assume that an  effective   $\theta (\vec{x}, t)_{ind}\neq 0$  is induced on a large scale as a result of collision~\cite{Kharzeev:2007tn}. The induced $\la\theta (\vec{x}, t)_{ind}\ra\neq 0$ obviously    violates local ${\cal{P}}$  and ${\cal{CP}}$ symmetries on the same scales $\mathbb L$ where $\theta (\vec{x}, t)_{ind}\neq 0$  is  correlated. It  may  generate  a number of  ${\cal{P}}$  and ${\cal{CP}}$ violating effects, such as Chiral Magnetic Effect (CME).  One of the critical questions for the applications of the CME to heavy ion collisions is  a correlation length of the induced $\la\theta (\vec{x}, t)_{ind}\ra\neq 0$.  Why are these  ${\cal{P}}$  odd domains     large?
       
          We speculate \cite{Zhitnitsky:2012im} that the crucial   element in understanding this key question is deeply rooted to the properties of long range  order  advocated in this work. 
        The   ${\cal{P}}$  odd domains are identified with coherent ${\cal{P}}$  odd sheets making the  ``skeleton" from section \ref{skeleton}.  This long range order  may explain why CME is  operational in this system and how the asymmetry can be coherently accumulated from entire system.    This identification  would  justify   the effective Lagrangian  approach advocated in \cite{Kharzeev:2007tn}  when   $\theta (\vec{x}, t)_{ind} $ is treated as slow background field 
      with  correlation length much larger than any conventional QCD fluctuations, $\mathbb L\gg \Lqcd^{-1}$.   Some of the related questions on CME can   in fact be tested  in deformed QCD~\cite{Zhitnitsky:2012ej}.
       
       3) The key element   advocated in the present work is the emergence  of long range pure topological fields (such as $a(\mathbf{x}), b(\mathbf{x})$ fields  related to the Veneziano ghost as discussed in section  \ref{BF-section}).  These auxiliary fields do not have kinetic terms, they do  not propagate, they do not violate unitarity or causality, but nevertheless they  do contribute to some local characteristics of the system such as the topological susceptibility and $\theta$ dependent portion of energy associated with it. The unique features of these topological fields have  inspired a proposal \cite{Zhitnitsky:2011tr,Zhitnitsky:2011aa,Zhitnitsky:2012im} that the observed dark energy in the universe may in fact be related to such   pure topological, non- propagating, long- ranged degrees of freedom\footnote{This idea  should be  contrasted with conventional approaches in study  of the   nature of dark energy  when one normally introduces   some real physical propagating dynamical degrees of freedom  into the system with highly  tuned parameters to match the observations.  }.
       We refer  to the original papers  \cite{Zhitnitsky:2011tr,Zhitnitsky:2011aa,Zhitnitsky:2012im}  for the details on this idea  as well as for a large number of   references where this proposal was (successfully)  confronted  with current observational data.

   Finally, one should note that the crucial physics which is responsible for a number of striking features discussed in this paper 
   and which are due to the tunnelling events between ``degenerate" winding states  can be in principle simulated   in a laboratory.
       To be more specific, when  the Maxwell system is defined on a compact manifold there will be a
       fundamentally new long range contribution to the vacuum energy ~\cite{Cao:2013na}. 
    This  extra contribution to the Casimir pressure is not related to the physical   propagating photons with two transverse polarizations, similar to our discussions of the topological susceptibility.  
  This  novel type of the vacuum energy is 
       very similar in nature and in spirit 
   to the effects discussed in the present work, and can  hopefully   be tested in a laboratory.

   \section*{Acknowledgements}
  I am    thankful to    Dima Kharzeev,   Pietro Faccioli, Edward Shuryak  and other  participants of 
 the workshop ``P-and CP-odd effects in hot and dense matter", Brookhaven, June, 2012, for     discussions on   long range order in QCD and its possible manifestations. 
 I am also thankful to Erich Poppitz,  Mithat  Unsal and other  participants of 
 the workshop ``Continuous Advances in QCD", Minneapolis, May, 2013, where this work has been presented, for useful and stimulating discussions.  
 I am also thankful to Misha Polikarpov and Pavel Buividovich for correspondence   on  feasibility to study  some long range effects in lattice QCD.  
 I am also thankful to I.Horvath and  G. Volovik for correspondence and useful comments.  Finally,  it is a great pleasure for me to thank my former student Max Metlitski for   chatting    on possible applications  of the ideas discussed   in this work to    CM systems.
 This research was supported in part by the Natural Sciences and Engineering Research Council of Canada.

\end{document}